\documentclass{aa}
\usepackage[latin1]{inputenc}
\usepackage{graphicx,psfig}
\usepackage{amssymb}
\usepackage{mathrsfs}
\usepackage{natbib}
\bibpunct{(}{)}{;}{a}{}{,}
\usepackage{txfonts}

\newcommand\kms{\ensuremath{\mbox{km}\,\mbox{s}^{-1}}}
\newcommand\vsini{\ensuremath{v\sin i}}
\newcommand\Msol{\ensuremath{\mathscr{M}_\odot}}
\newcommand\td{\hbox{$^{\times}\!\!/$}}

\begin{document}

\title{Rotational velocities of A-type stars\thanks{Table~\ref{list} is only available in electronic form at the CDS via anonymous ftp to {\tt cdsarc.u-strasbg.fr (130.79.125.5)} or via {\tt http://cdsweb.u-strabg.fr/Abstract.html}}}

\subtitle{III. Velocity distributions}

\author{F. Royer\inst{1}
   \and J. Zorec\inst{2}
   \and A.~E. G\'omez\inst{1}
}

\offprints{Frédéric Royer,\\ \email{frederic.royer@obspm.fr}}
 
\institute{GEPI/CNRS UMR 8111, Observatoire de Paris, 5 place Jules Janssen, 92195 Meudon cedex, France
      \and CNRS, Institut d'Astrophysique de Paris, 98 bis boulevard Arago, 75014 Paris, France
}

\date{Received  / Accepted }

\titlerunning{Rotational velocities of A-type stars. III.}

\abstract
{}
{In this work, a sample of \vsini\ of B9 to F2-type main sequence single stars has been built from highly homogeneous \vsini\ parameters determined for a large sample cleansed from objects presenting the Am and Ap phenomenon as well as from all known binaries. The aim is to study the distributions of rotational velocities in the mass range of A-type stars for normal single objects.}
{Robust statistical methods are used to rectify the \vsini\ distributions from the projection effect and the error distribution. The equatorial velocity distributions are obtained for an amount of about $1100$ stars divided in six groups defined by the spectral type, under the assumption of randomly orientated rotational axes.}
{We show that late B and early A-type main-sequence stars have genuine bimodal distributions of true equatorial rotational velocities due probably to phenomena of angular momentum loss and redistribution the star underwent before reaching the main sequence. A striking lack of slow rotators is noticed among intermediate and late A-type stars. The bimodal-like shape of their true equatorial rotational velocity distributions could be due to evolutionary effects.}
{}


\maketitle


\section{Introduction}
 The observed rotation of stars undergoing the early main sequence (MS) evolutionary stages, known as the ``dwarf'' stage, is the result of a long process of global angular momentum loss and angular momentum redistribution inside the stars. This process begins with the stellar formation phases, which encompass the fragmentation of rotating clouds \citep{Bor81,Mao_97} and the angular momentum losses required to solve the {\em angular momentum problem} through the initial magnetic braking \citep{MosMon85b,MosMon85a} and bipolar overflows \citep{Puz85}. They give the stars the characteristics that can be considered as their ``initial conditions'' at the birth-line. Further pre-main sequence (PMS) processes of angular momentum exchange and/or magnetic locking with the accretion disc \citep{Kol91,Eds_93}, angular momentum losses by stellar winds \citep{Scn62,Shu_00} and internal angular momentum redistribution phenomena, shape the internal structural patterns and rotational characteristics of stars at the zero-age main sequence (ZAMS) that may be called the ``initial conditions'' of stars in the MS.

 Depending on the stellar internal structural pattern thus acquired, the chemical composition and the morphology of the magnetic fields, fossil and those created in the star \citep{Spt99,Spt02}, the radiative and convective regions can be more or less coupled hydrodynamically \citep{Lar_81} and/or magnetically \citep{Bas03} to produce the internal angular momentum profile \citep{EnlSoa81,Str_84,Pit_90,MarBrr91,Kes_95,Som_93,Som_01}.

Thus, according to differences in the processes and respective efficiencies to produce loss and exchange of angular momentum with the environment, as well as to many ways the stars can manage to redistribute the angular momentum in their interior, a variety of internal rotational laws in the ZAMS among objects with the same mass can be present. If only rigid rotation were prevalent, rather simple functions would characterize the probability distribution of a single quantity: the angular velocity rate $\Omega/\Omega_\mathrm{c}$ ($\Omega_\mathrm{c}$ is the critical rotational velocity). Such distributions would probably be unimodal. On the other hand, the conjunction of a variety of rotational laws and rotational rates of coeval stars with the same mass in a single rotational distribution can have a multimodal aspect.

 \citet{Gue82} found that late B-type stars in clusters have bimodal rotational velocity distributions, while they are unimodal for the same class of field stars. Bimodality was also observed for the rotation of young solar mass stars in Orion \citep{AteHet92,ChiHet96,Het_01,Bas03}. For masses intermediate to the above cases, \citet{AbtMol95} found bimodal-like distributions among A0--F0 objects in the MS, where the component owing to low rotators was ascribed to the chemically peculiar Am and Ap stars.

However, a high fraction of objects with the Am phenomenon were found to be binaries \citep{Dei00}. Regarding Ap stars, their peculiarity requires a given time after the ZAMS to appear and thus, it can be considered somehow evolution-dependent \citep{Hug_00a,Stn00}. All these quoted facts can then be summarized as follows:
\begin{enumerate}
\item Although it is known that the chemical separation in Am and Ap stars are not produced by the slow rotation, slow rotation favors their appearance, so that both are correlated. In Am stars slow rotation can be due to their binary character through tidal braking, while in Ap stars could be caused in part by magnetic braking;
\item Since in the whole mass interval $1.3 \la \mathscr{M}/\Msol \la 3$ bimodality is clearly apparent only in cluster late B-type stars, it may correspond to some star formation characteristics.
\end{enumerate}
It is tempting then to review anew the velocity distributions in the main sequence phase (``dwarf'' evolutionary state) of stars in the mentioned mass range, in order to detect possible signatures on differences in stellar formation characteristics. It is not excluded, however, that some particular signatures can be present in the rotational velocity distributions of those A-type stars that make the transition between objects with and without convectively unstable envelopes. These last can likely be due to stellar structure properties rather than to formation circumstances.

 Thus, the aim of this paper is to see: 1) whether single late B- and A-type field stars in the early main sequence evolutionary phases have unimodal or multimodal rotational distributions; 2) whether the rotational distributions of intermediate and late A-type stars bear signatures related with the complexity of their stellar envelope/atmospheric structure.

 The data used in this paper and the selection of samples are described in Sect.~\ref{Sect_data}. The \vsini\ distributions are presented in Sect.~\ref{Sect_vsinidist}. Section~\ref{Sect_vdist} gives details on the statistical processing of stellar samples to get the equatorial velocity distributions from the observed \vsini\ values and describes the resulting distributions. Finally, the results are discussed in Sect.~\ref{Sect_Disc} and summarized in Sect.~\ref{Sect_Conc}.


\section{Rotational velocities data}
\label{Sect_data}
\subsection{Main \vsini\ data sources}

 When studying distributions of stellar parameters, not only large samples, but also as homogeneous as possible are suited to avoid or minimize possible biases. The hitherto largest list of rotational velocities is provided by \citet{GliSti00}, who gathered \vsini\ parameters from the literature of over 11\,000 stars of all spectral types without magnitude limitations. Nonetheless, this catalog is built up from about 200 different bibliographic sources and it can be therefore highly heterogeneous. We preferred then to base our study on two homogeneous data sets: on one hand, the \vsini\ obtained using the Fourier transform method (FT) by \citet[ hereafter Papers I and II, or I~$\cup$~II]{Ror_02a,Ror_02b} and on the other hand, the \vsini\ determined by \citet[ hereafter called AM]{AbtMol95} and \citet[ hereafter called ALG, or AMALG when both papers are referred to]{Abt_02} from the FWHM of \ion{He}{i} and \ion{Mg}{ii} lines calibrated into the \vsini\ scale of \citet{Slk_75}. Nevertheless, these two data sets are different. The AMALG sample is complete down to the apparent magnitude $V=6.5$ mag and contains A- and B-type stars from the northern hemisphere of all classes. 
The I~$\cup$~II sample concerns objects from both hemispheres, some down to magnitude $V=8$ mag, and observation of those with missing \vsini\ parameters were privileged. The selection criteria used to establish the I~$\cup$~II stellar sample are summed up in \citet{GeiMar_89}. 
\begin{figure}[!htp]
  \centering \resizebox{\hsize}{!}{\includegraphics{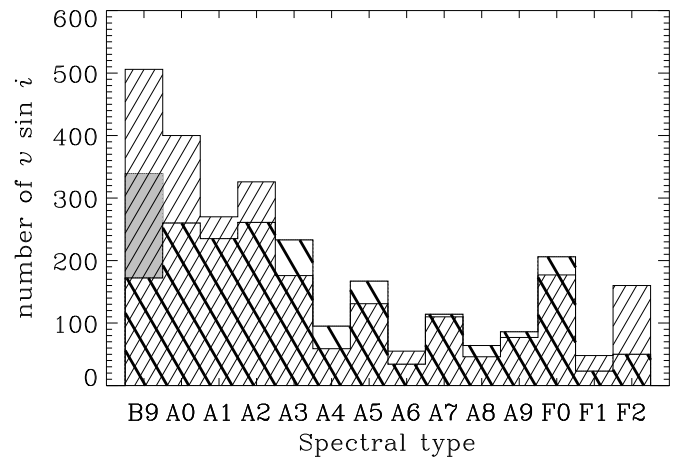}}
        \caption{Distribution of spectral types of the bright stars ($V<8$) with known \vsini\ in the spectral type range B9--F2. The {\bf thin-dashed} histogram represents the stars from \citet{GliSti00} and the {\bf thick-dashed} histogram is the sample from \citet{Ror_02b}. The gray histogram is for B9-type stars in \citet{Abt_02}.}
\label{histST}
\end{figure}

\begin{table*}[!htp]
\begin{center}
\caption{Highlight on the four common stars between \citet[
  SCBWP]{Slk_75}, \citet{Abt_02} and Papers~I and II (standard deviation is
  indicated; dash ``--'' stands for only one measurement) and comparison
  with data from the literature. \vsini\  from the literature are classified in three
  subgroups according to the way they are derived: by-product of a
  spectrum synthesis or frequency analysis of the lines profiles.
  Flags from HIPPARCOS catalog are indicated:
  variability flag H52 (C: constant, D: duplicity-induced variability,
  M: possibly micro-variable, P: periodic variable, --: no certain
  classification) and double annex flag H59 (C: component solution, --: no entry in the Double and Multiple Systems Annex). }
\label{vsini_litt}
\begin{tabular}{lrlrr@{\hspace*{1mm}}llllcc}
\hline
\hline
\multicolumn{1}{c}{Name} & \multicolumn{1}{c}{HD} &
\multicolumn{1}{c}{Sp. type} &  \multicolumn{6}{c}{\ensuremath{v\sin i}\ (\ensuremath{\mathrm{km}\,\mathrm{s}^{-1}})}
& \multicolumn{2}{c}{{\sc hipparcos}}        \cr
   &    &      &  {\sc scbwp} & \multicolumn{2}{c}{I $\cup$ II}&\multicolumn{3}{c}{\hbox{\raisebox{0.3em}{\vrule depth 0pt
         height 0.4pt width 1.5cm} literature \raisebox{0.3em}{\vrule
         depth 0pt height 0.4pt width 1.5cm}}} & H52 & H59\cr
   &     &      &    &    &      & \multicolumn{1}{c}{spec. synth.}& \multicolumn{1}{c}{freq. analysis}\cr
\hline
$\beta$~Tau   &  35497 & B5IV    & $60$ & $64$ &${\scriptstyle\pm 5}$&&&& --&--   \\
$\tau$~Her    & 147394 & B9III   & $30$ & $46$ &${\scriptstyle\pm 2}$&$32^{(1)}$&$32^{(2)}$&& P &--   \\
$\gamma$~Lyr  & 176437 & B9.5III & $60$ & $72$ &${\scriptstyle\pm 2}$&&&& M &--   \\
$\omega^2$~Aqr& 222661 & B9V   &$120$ &$150$ & --&&&& C &--   \\
\hline
\end{tabular}
\end{center}
\begin{tabular}{llll}
$^{(1)}$ \citet{Adn88} & $^{(2)}$ \citet{SmhDwy93}   \\
\end{tabular}
\end{table*}

The merging of \vsini\ data from the I~$\cup$~II sample and AM has already been dealt with in Paper~II. The distribution of the number of stars with measured \vsini\ in the resulting sample is displayed in Fig.~\ref{histST} (thick-dashed histogram). As a comparison, the same distribution for G\l\c{e}bocki \& Stawikowski's in the spectral type range B9--F2 is over-plotted (thin-dashed histogram).
In this figure we see that the exhaustive compilation by G\l\c{e}bocki \& Stawikowski contains a significantly higher amount of data for stars in the B9 to A0-type spectral interval and for F2-type stars.
 The lack noted of late B-type stars in I~$\cup$~II can, however, be nicely completed with the ALG data to obtain a merged sample that becomes comparable in size with the exhaustive collection in \citet{GliSti00} in the whole spectral range from B9 to F2 as shown by the gray shading in Fig.~\ref{histST}. So, the rotational velocity distributions studied in the present paper are based entirely on the I~$\cup$~II plus AMALG merged and homogenized sample of \vsini\ parameters.

\subsection{Merging with \citet{Abt_02} and the \vsini\ scale}
\label{merge}

\begin{figure}[!thp]
\resizebox{\hsize}{!}{\includegraphics{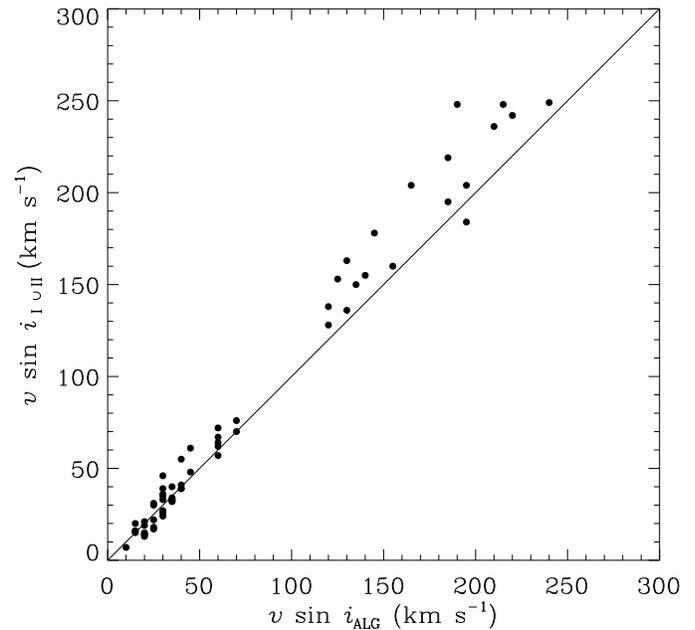}}
\hfill
\caption{Comparison of \vsini\ data for the 64 common stars from Papers I and II (I $\cup$ II) and \citet[ ALG]{Abt_02}. The solid line stands for the first bisector. }
\label{comp-vsini1}
\end{figure}

 AMALG determined \vsini\ for nearly 2800 B- and A-type stars in the northern hemisphere using the full width at half maximum ($FWHM$) of the \ion{He}{i} 4471 and \ion{Mg}{ii} 4481 \AA\ lines calibrated in the \vsini\ scale of \citet{Slk_75}.
The merging with the A-type part of the sample (AM) was carried out in the previous paper.
The intersection of the B-type part (ALG) with I~$\cup$~II contains 64 common stars, essentially of spectral types B8, B9 and B9.5. The comparison of both \vsini\ scales is displayed in Fig.~\ref{comp-vsini1}. From this diagram we can see that there is a systematic deviation, in the sense that the $\vsini\ga 100$\,\kms\ from I~$\cup$~II are on average $\approx16$\,\% larger than those in ALG. This confirms the global tendency of \vsini\ being lower when determined with the Slettebak et al. calibration \citep[cf.][]{Ror_02b}. Both scales coincide, however, for $\vsini\la 80$\,\kms. The linear regression line between both scales was determined with GaussFit \citep{Jes_98a,Jes_98b}, a robust least squares minimization program to obtain empirical functions:
\begin{equation}
\label{III-ALG}
\vsini_{\mathrm{I}\cup\mathrm{II}} = 1.21{\scriptstyle\pm 0.04}\,\vsini_\mathrm{ALG}-6.1{\scriptstyle\pm 1.3}.
\end{equation}

In papers I~$\cup$~II there are also four standard stars common with Slettebak et al. which are listed in Table~\ref{vsini_litt} and whose \vsini\ can be compared. Other details on these objects are given in Appendix~\ref{Notes1}. 

 It is worth emphasizing that Eq.~\ref{III-ALG} was used only for stars of late B spectral types, i.e. B9 and B9.5. The resulting I~$\cup$~II plus AMALG sample contains thus 2167 \vsini\ parameters of stars from spectral type B9 to F2.

 Let us comment briefly the reason of aligning the \vsini\ parameters to our I~$\cup$~II rotational velocity scale. 
According to the gravity darkening (GD) effect, the fastest rotating regions (equatorial zone) contribute less to the global line flux than higher latitudes, in particular when it concerns \ion{He}{i} 4471 and \ion{Mg}{ii} 4481\,\AA\ lines. It has been shown \citep[cf.][]{Sty68b,Frt_05a} that the interpretation of spectral lines of fast rotating stars with classical methods lead systematically to underestimated values of \vsini\ . Nevertheless, \citet{Slk_75}, who take GD into account, report that their \vsini\ parameters are systematically lower than those issued from classical methods by roughly 15\,\% among early stars up to spectral type A2, which is contrary to what is predicted by modern calculations of non-LTE rotationally broadened lines affected by GD \citep{Frt_05a}. The difference noted by \citet{Slk_75} drops to only some percents for stars later than A2. \citet{Hoh04} has also re-examined recently Slettebak's et al. \vsini\ scale and confirmed the above mentioned underestimation.

  Since the \vsini\ parameters in papers I~$\cup$~II are systematically higher than those by AMALG, which mirror Slettebak's et al. (1975) scale, we adopted a conservative position and decided to use our FT scale of \vsini\ for the merged I~$\cup$~II plus AMALG data source. Except for the assumptions that underlie the FT method, the I~$\cup$~II \vsini\ scale is independent from any other calibration. This scale has been moreover confirmed recently by a new more robust FT method \citep{ResRor04a} and proved to be consistent also with new \vsini\ determinations \citep[e.g.][]{ErrNoh03,Fel03}.

\subsection{Sample definition}\label{sampd}

\subsubsection{Luminosity classes}

We attempt at studying the characteristics of the \vsini\ distributions in the MS evolutionary phase. Since we cannot determine ages for all studied stars, we proceed to a statistical selection of those objects that could be meant to be in the MS stage. To this end, we make a selection of ``dwarf'' stars and retain only those with luminosity classes ranging from V to IV. This luminosity class criterion is large enough to take into account possible rotationally induced effect in fast rotators, which make stars look more luminous or evolved than they really are \citep{CosSon77,Cos_91,Frt_05a}.
The luminosity classes used for this selection are the ones compiled in the HIPPARCOS catalog \citep{Hip}.

\subsubsection{Spectral types}

\begin{figure}[!htp]
\resizebox{\hsize}{!}{\includegraphics{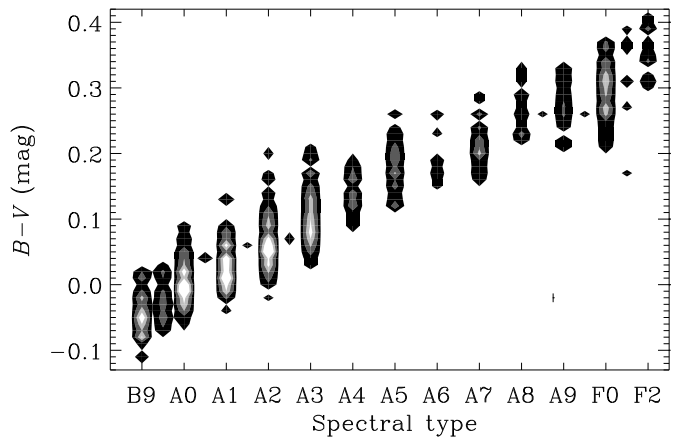}}
\hfill
\caption{Comparison between the spectral class and the $B-V$ color index (both from HIPPARCOS catalog) for the stars in the \vsini\ sample, with luminosity classes V, IV-V and IV. The colored contour indicates the number of stars per bin of $0.01$\,mag, per spectral class: black --- $1$ to $5$ stars; dark gray --- $5$ to $10$ stars; light gray --- $10$ to $15$ stars; white --- more than $15$ stars. The average error bar in $B-V$ is indicated: $\bar{\sigma}_{B-V}=0.0067$\,mag.}
\label{TS}
\end{figure}

  The selected 1500 stars were finally divided into spectral type subsamples,
based on the spectral types compiled in the HIPPARCOS catalog.
In establishing the subgroups, we looked for a compromise between the sample size and the radii dependence with mass. The correspondence between spectral type and radius used is from \citet{HasHee81}. The relation between spectral class and $B-V$ color index is shown in Fig.~\ref{TS}. The mean error on $B-V$, $\bar{\sigma}_{B-V}= 0.0067$\,mag, is small in comparison with the color overlap between the different spectral classes. A running standard deviation (on 200 points) on this relation enables us to estimate the error committed on the spectral classification: about one sub-class for $B-V\la 0.15$\,mag, and about two sub-classes for $B-V\ga 0.15$\,mag. The color index $B-V=0.15$\,mag approximately corresponds to the spectral class A5 \citep{Coy_70}. Color indices can be rather strongly affected in late type stars by fast rotation effects if rotational rates are $\Omega/\Omega_{\rm c}\ga 0.9$ \citep{CosSon77,Cos_91,Frt_05a}. Since we cannot correct in advance individual spectral types for possible induced rotational changes, we shall conclude a posteriori that this phenomenon may in principle be affecting only a low fraction of stars (see Fig.~\ref{omegaprob}:
$7.6$\,\% in B9, $5.4$\,\% in A0--A1, $4.2$\,\% in A2--A3, $4.2$\,\% in A4--A6, $3.6$\,\% in A7--A9 and $2.5$\,\% in F0--F2).
The full spectral range B9--F2 was thus divided into six groups, which are listed in Table~\ref{subsamples}.

\begin{table}[htbp]
  \begin{center}
\caption{Subsamples of Main-Sequence stars (classes V, IV-V and IV):
  the mean radii ${\overline R}/R_\odot$ and the mean masses ${\overline\mathscr{M}}/\Msol$ are derived from
  \citet{HasHee81}, as well as the relative variation of the radius $\Delta R$ and of the mass $\Delta\mathscr{M}$
  along the interval of spectral type. For each subsample, the number of stars is given.}
\label{subsamples}
\setlength\tabcolsep{2pt}
    \begin{tabular}{cccccc}
\hline
\hline
Spec. types & ${\overline R}/R_\odot$ & $\Delta R$ (\%) & ${\overline\mathscr{M}}/\Msol$ & $\Delta\mathscr{M}$ (\%) &\# stars \\
\hline
       B9     & $1.95$ &      --    & $2.34$ &      --    & $198$ \\
       A0--A1 & $1.79$ & $\pm  0.8$ & $2.04$ & $\pm  1.5$ & $397$ \\ 
       A2--A3 & $1.70$ & $\pm  1.2$ & $1.90$ & $\pm  2.1$ & $377$ \\ 
       A4--A6 & $1.60$ & $\pm  2.5$ & $1.73$ & $\pm  3.8$ & $182$ \\ 
       A7--A9 & $1.48$ & $\pm  2.7$ & $1.52$ & $\pm  4.3$ & $190$ \\ 
       F0--F2 & $1.37$ & $\pm  2.2$ & $1.37$ & $\pm  2.2$ & $197$ \\ 

\hline
    \end{tabular} 
  \end{center}
\end{table}

\begin{figure}[!htp]
\resizebox{\hsize}{!}{\includegraphics{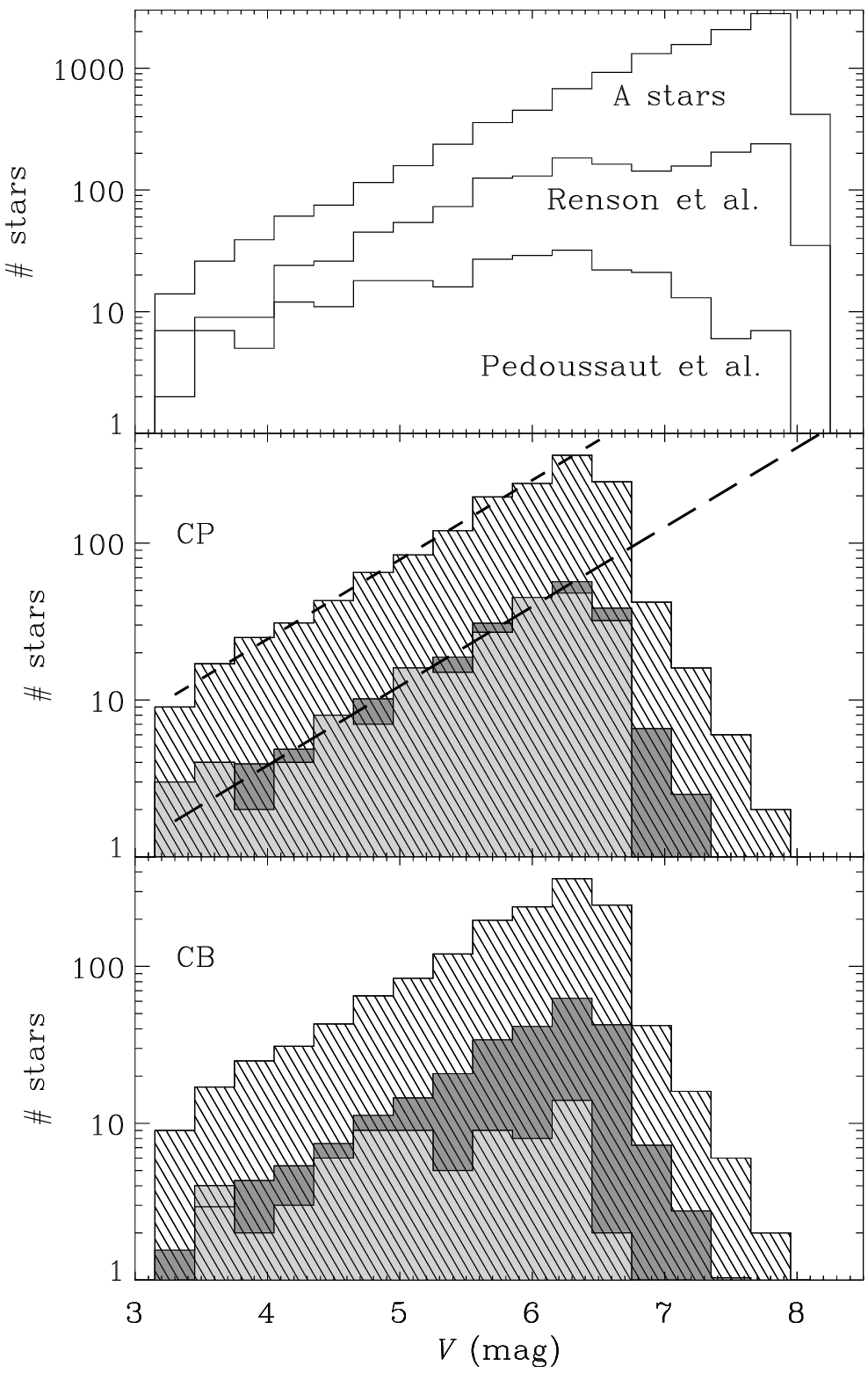}}
\hfill
\caption{Distributions of $V$ magnitudes and completeness of the CP and CB stars. The numbers of stars are given in logarithmic scale. {\bf Top panel)} the
  distributions (in the range $3$--$8$\,mag) are displayed for the A-type stars in the HIPPARCOS catalog, and their intersection with the lists given by \citet{Ren_91} and \citet{Pet_85}.
{\bf Middle  panel)} the distributions are shown for the \vsini\ sample (hatched histogram) and the selected CP stars (light gray). Dark gray histogram is the \vsini\ sample scaled to the number of CP stars in the $V$-magnitude range $4$--$6$\,mag.
{\bf Bottom  panel)} the distributions are shown for the \vsini\ sample (hatched histogram) and the selected CB stars from the list by \citet{Pet_85} only (light gray). Dark gray histogram is the \vsini\ sample scaled to the number of binary stars in the $V$-magnitude range $4$--$5$\,mag.
  }
\label{completV}
\end{figure}

\subsubsection{Chemically peculiar and binary stars}

In order to clean the sample from stars whose rotation could have been modified by tidal or magnetic braking, chemically peculiar stars (CP) and ``close'' binary stars (CB) have been {\bf separated from ``normal'' stars}. This choice is moreover justified because of the a priori selection biases in the I~$\cup$~II sample. As regards the southern sample from Paper~I, selection criteria are described by \citet{GeiMar_89}. They focused on ``normal'' A-type stars, excluding any known peculiar or spectroscopic binary stars. The same criteria have been used for the northern sample in Paper~II.

\paragraph{CP stars} (chemically peculiar stars):
The catalog of Ap and Am stars from \citet{Ren_91} is used to identify peculiar stars, as well as the spectral classification made by AM and ALG. The number of CP stars in each subsample is summarized in Table~\ref{moments}. 

\paragraph{CB stars} (``close'' binary stars):
The tidal effect in a multiple  system tends to synchronize the axial rotation period on the orbital period. This category of stars was selected on criteria based on HIPPARCOS data and spectroscopic data. Except for a few of them, all selected stars are in the HIPPARCOS catalog \citep{Hip}. The binaries detected by the satellite with $\Delta Hp<4$\,mag are flagged as CB stars. The compilation by \citet{Pet_85} of spectroscopic binaries is used to complete the identification, together with the ``Eighth catalog of the orbital elements of Spectroscopic Binaries'' \citep{Ban_89}.

 In what follows, all stars that do not obey to CB or CP criteria are simply called ``normal''. Since these ``normal'' stars constitute the bulk of our stellar sample, it is worth having a statistical inference of the fraction of stars where the CP and CB characters are still unknown.

\paragraph{Completeness of the knowledge of CP stars:}
 a large part of the data sample was observed by Abt (AMALG), who derived at the same time the \vsini\ values, determined the spectral classification and identified the chemical peculiarities. In addition, the catalog of \citet{Ren_91} was also used to account for peculiar stars. As a reference to test the completeness of this catalog we plot in the top panel of Fig.~\ref{completV} all A-type stars in the HIPPARCOS catalog, the CP stars in the \citet{Ren_91} catalog and P\'edoussaut's et al. (1985) compilation of spectroscopic binaries. The number of stars is displayed in logarithmic scale against the $V$-magnitude. In the middle panel of Fig.~\ref{completV} there is the merged I~$\cup$~II plus AMALG data set (hatched histogram), where the short-dashed line is simply a fit of their distribution up to $V=6.5$ mag. The \emph{light gray} histogram gives the CP stars identified in this paper over which we have superimposed the long-dashed linear fit valid up to $V=6.5$ mag. We see that the fraction of CP stars to all stars in the merged sample is fairly constant for all magnitudes brighter than $V=6.5$ mag and it represents roughly 15\,\% of objects. The same fraction, and, in the same magnitude interval, is also seen in the top panel. Making the hypothesis that the proportion of CP stars in our merged ensemble remains at 15\,\% also for $V\ga6.5$ mag, the scaled distribution (dark gray) shows the deficit of undetected CP, which amounts only to some 20 stars.

\paragraph{Completeness of the knowledge of CB stars:}
 the identification of the binary nature of stars is a difficult problem and can lead to partially satisfying solutions. Most of the time it requires surveys with spectroscopic monitoring. The use of HIPPARCOS results to account for close binary stars simply does not allow detection of the closest ones \citep[$\rho<0\farcs 1$--$0\farcs15$,][]{Hip} or those with large magnitude difference ($\Delta Hp>3.5$--4\,mag). Considering the mass--luminosity relation from \citet{Gei_99}, the magnitude difference leads approximately to $\Delta Hp\approx-2.51\,\Delta(\log\mathscr{L})\approx-11.42\,\Delta(\log\mathscr{M})$ and then for $\Delta Hp =4$\,mag, the corresponding mass ratio is $q\approx 0.45$.

 \citet{Pet_85} provide an ``exhaustive'' list of spectroscopic binaries and the estimation of the lack of detected binaries is based on this list only. As it is obvious, we aim at removing binaries from the \vsini\ sample to minimize rotational changes induced by tidal braking and the consequent synchronization. The completeness limit magnitude of the list of spectroscopic binaries is estimated in the top panel in Fig.~\ref{completV}, which hardly reaches $V=5$\,mag. The distribution of the \vsini\ sample is scaled as in the previous paragraph, according to the number of spectroscopic binaries in the magnitude range $4$--$5$\,mag. The scaled distribution (dark gray, bottom panel) gives an estimate of possible undetected spectroscopic binaries, which amounts to some $200$ stars, mainly for $V\ga5$ mag.  It is however worth noting that the frequency of binaries decreases with decreasing mass ratio $q$ \citep{Wof78} and that correspondingly the synchronization time increases \citep{Zan75}. On the other hand, the maximum of the period distribution of spectroscopic binaries in the spectral range A0--A9 is for $P\approx5$\,d and that for $P\approx1$\,d the frequency slumps to $25$\,\% \citep{Koe81}. The most substantial fraction of stars in the A0--A9 range with \vsini\ that can be expected to be affected by tidal braking must then correspond to $q\approx0.5$ and $P\approx2$\,d. The synchronization time is a strong function of the $D/R$ ratio, where $D$ is the separation of components and $R$ the radius of the primary: $t_{\rm syn}\propto(D/R)^{8.5}$ \citep{Zan75}. For $q\approx0.5$, $P\approx2$\,d, average stellar mass $1.8\,\Msol$ and radius $1.6\,R_{\odot}$ we obtain $D/R\la6$, so that $t_{\rm syn}\approx 0.5\,t_\mathrm{MS}$, where $t_\mathrm{MS}$ is the time spent in the main sequence. At shorter periods the number of spectroscopic stars decreases, while for longer periods the synchronization time increases strongly, i.e. for $P\approx5$\,d, it is $t_{\rm syn}\approx 100\,t_\mathrm{MS}$. This means that the possibly $13$\,\% unidentified binaries cannot blur sensitively our statistics. Moreover, it is easy to show that a slowing down of rotation by $50$\,\% of the rotational velocities in the missing binaries would affect the overall average estimate of \vsini\ by less than $4$\,\%, because more than half have $P\ga5$\,d.
 
\paragraph{}It is worth noting that these deficits of stars are not accountable only to slow rotators. Among the 218 identified ``CP'' stars, only 87 have $\vsini<70$\,\kms, i.e. about $40$\,\%. As far as the spectroscopic binaries are concerned, among the 79 stars in common with the list by \citet{Pet_85}, 43 stars have $\vsini<70$\,\kms, i.e. about $54$\,\%.

\paragraph{}The number of objects in the different spectral-type subgroups, and the different categories (``normal'', CB, CP) are given in Table~\ref{moments}. All the objects are listed in Table~\ref{list}.


\section{Projected rotational velocities distributions}
\label{Sect_vsinidist}

\begin{figure*}[!htp]
\resizebox{\hsize}{!}{\includegraphics{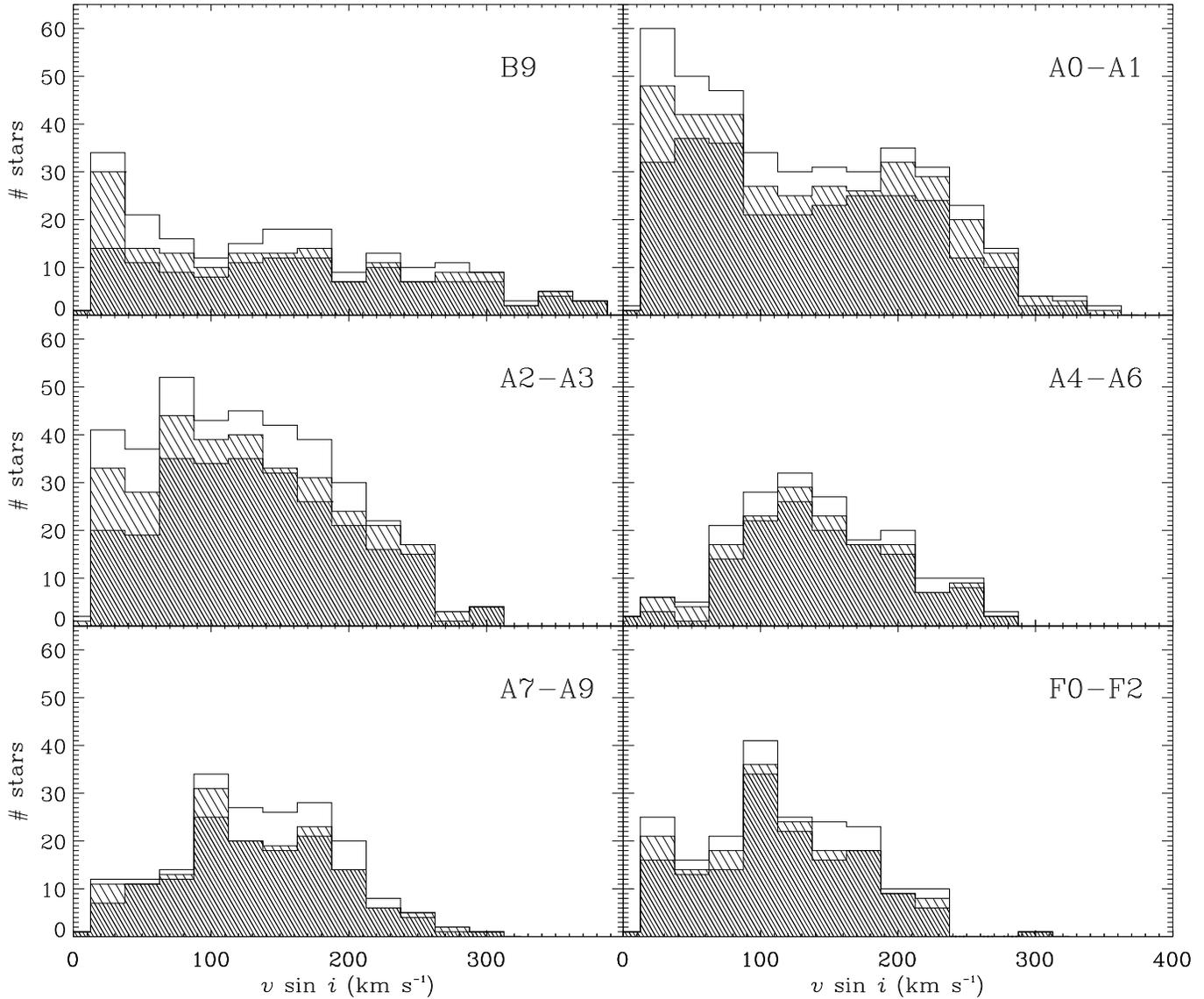}}
\hfill
\caption{Cumulative distributions of \vsini\ of the six subsamples of main
  sequence stars from B9- to F2-type. Dark hatched, light hatched and plain white
  areas, from bottom to top, respectively stand for ``normal'', CP and CB stars.}
\label{distrib1}
\end{figure*}

The histograms of \vsini\ of each of the six subsamples are displayed
in Fig.~\ref{distrib1}. The associated statistical estimators are
listed in Table~\ref{moments}. Distinction has been made between the ``total'' subsample ans its components:
``normal'', ``CP'' and ``CB'' stars (respectively represented as dark hatched, light hatched and
plain white areas in Fig.~\ref{distrib1}). The distinction in Table~\ref{moments} is also made according to the luminosity class: ``all''  classes (V, IV-V and IV), ``V'' only and ``IV'' only.
Mean, median and dispersion estimators are also given for each group.

\begin{table}[!htbp]
  \begin{center}
\caption{Moments of the \vsini\ distributions for each spectral
  subsample. ``total'' corresponds to the whole subsample (including CP and CB stars);  ``normal'', CP and CB stars are also distinguished. For both ``total'' and ``normal'' subsamples, the moments are given for different groups of luminosity class: all (IV, IV-V, V), V and IV {\bf (class IV-V luminosity stars only are not considered because they represent a very small fraction of the sample)}. The number of stars (\#), the center and the dispersion of the distribution are given. Centers are estimated using the median and the mean, and the dispersion is evaluated with the standard deviation.}
\label{moments}
\setlength\tabcolsep{4pt}
\begin{tabular}{llrrr@{\hspace*{1mm}}lr@{\hspace*{1mm}}lr@{\hspace*{1mm}}l}
\hline
\hline
\multicolumn{3}{c}{Subsample} & \# & \multicolumn{4}{c}{Center
  (\kms)}   & \multicolumn{2}{c}{Disp.}\\ 
      &        &               &         & \multicolumn{2}{c}{median} & \multicolumn{2}{c}{mean}  &
              \multicolumn{2}{c}{(\kms)} \\
\hline
       &      & all & $198$ & $137$ & ${\scriptstyle\pm  8.8}$ &  $145$ & ${\scriptstyle\pm  7.0}$&  $ 98$ & ${\scriptstyle\pm  4.9}$\\ 
       & total&  V  & $163$ & $153$ & ${\scriptstyle\pm  9.6}$ &  $154$ & ${\scriptstyle\pm  7.7}$&  $ 98$ & ${\scriptstyle\pm  5.4}$\\ 
       &      & IV  & $ 24$ & $ 68$ & ${\scriptstyle\pm 23.9}$ &  $102$ & ${\scriptstyle\pm 19.1}$&  $ 93$ & ${\scriptstyle\pm 13.5}$\\ 
\cline{2-10}
B9     &      & all & $125$ & $159$ & ${\scriptstyle\pm 11.0}$ &  $161$ & ${\scriptstyle\pm  8.7}$&  $ 98$ & ${\scriptstyle\pm  6.2}$\\ 
       &normal&   V & $108$ & $165$ & ${\scriptstyle\pm 11.8}$ &  $167$ & ${\scriptstyle\pm  9.4}$&  $ 98$ & ${\scriptstyle\pm  6.7}$\\ 
       &      &  IV &  $11$ & $103$ & ${\scriptstyle\pm 32.0}$ &  $110$ & ${\scriptstyle\pm 25.6}$&  $ 85$ & ${\scriptstyle\pm 18.1}$\\ 
\cline{2-10}
       & CP   &     & $ 36$ & $ 49$ & ${\scriptstyle\pm 20.4}$ &  $ 97$ & ${\scriptstyle\pm 16.3}$&  $ 98$ & ${\scriptstyle\pm 11.5}$\\ 
       & CB   &     & $ 37$ & $138$ & ${\scriptstyle\pm 17.6}$ &  $136$ & ${\scriptstyle\pm 14.1}$&  $ 86$ & ${\scriptstyle\pm 10.0}$\\ 
\hline
       &      & all & $397$ & $123$ & ${\scriptstyle\pm  5.2}$ &  $129$ & ${\scriptstyle\pm  4.1}$&  $ 82$ & ${\scriptstyle\pm  2.9}$\\ 
       & total&   V & $363$ & $123$ & ${\scriptstyle\pm  5.4}$ &  $131$ & ${\scriptstyle\pm  4.3}$&  $ 83$ & ${\scriptstyle\pm  3.1}$\\ 
       &      &  IV & $ 24$ & $ 80$ & ${\scriptstyle\pm 16.2}$ &  $ 88$ & ${\scriptstyle\pm 12.9}$&  $ 63$ & ${\scriptstyle\pm  9.1}$\\ 
\cline{2-10}
A0--A1 &      & all & $271$ & $128$ & ${\scriptstyle\pm  5.9}$ &  $131$ & ${\scriptstyle\pm  4.7}$&  $ 78$ & ${\scriptstyle\pm  3.4}$\\ 
       &normal&   V & $248$ & $128$ & ${\scriptstyle\pm  6.2}$ &  $131$ & ${\scriptstyle\pm  5.0}$&  $ 78$ & ${\scriptstyle\pm  3.5}$\\ 
       &      &  IV & $ 14$ & $100$ & ${\scriptstyle\pm 21.4}$ &  $105$ & ${\scriptstyle\pm 17.1}$&  $ 64$ & ${\scriptstyle\pm 12.1}$\\ 
\cline{2-10}
       & CP   &     & $ 69$ & $123$ & ${\scriptstyle\pm 14.5}$ &  $139$ & ${\scriptstyle\pm 11.6}$&  $ 96$ & ${\scriptstyle\pm  8.2}$\\ 
       & CB   &     & $ 57$ & $107$ & ${\scriptstyle\pm 13.9}$ &  $113$ & ${\scriptstyle\pm 11.1}$&  $ 84$ & ${\scriptstyle\pm  7.8}$\\ 
\hline
       &      & all & $377$ & $123$ & ${\scriptstyle\pm  4.5}$ &  $125$ & ${\scriptstyle\pm  3.6}$&  $ 69$ & ${\scriptstyle\pm  2.5}$\\ 
       & total&   V & $307$ & $131$ & ${\scriptstyle\pm  4.9}$ &  $133$ & ${\scriptstyle\pm  3.9}$&  $ 68$ & ${\scriptstyle\pm  2.7}$\\ 
       &      &  IV & $ 60$ & $ 70$ & ${\scriptstyle\pm  9.0}$ &  $ 80$ & ${\scriptstyle\pm  7.2}$&  $ 56$ & ${\scriptstyle\pm  5.1}$\\ 
\cline{2-10}
A2--A3 &      & all & $258$ & $128$ & ${\scriptstyle\pm  5.2}$ &  $133$ & ${\scriptstyle\pm  4.2}$&  $ 67$ & ${\scriptstyle\pm  3.0}$\\ 
       &normal&   V & $210$ & $133$ & ${\scriptstyle\pm  5.7}$ &  $140$ & ${\scriptstyle\pm  4.5}$&  $ 66$ & ${\scriptstyle\pm  3.2}$\\ 
       &      &  IV & $ 41$ & $ 71$ & ${\scriptstyle\pm 11.7}$ &  $ 93$ & ${\scriptstyle\pm  9.3}$&  $ 60$ & ${\scriptstyle\pm  6.6}$\\ 
\cline{2-10}
       & CP   &     & $ 60$ & $ 84$ & ${\scriptstyle\pm 12.5}$ &  $105$ & ${\scriptstyle\pm 10.0}$&  $ 77$ & ${\scriptstyle\pm  7.1}$\\ 
       & CB   &     & $ 59$ & $102$ & ${\scriptstyle\pm 10.3}$ &  $109$ & ${\scriptstyle\pm  8.2}$&  $ 63$ & ${\scriptstyle\pm  5.8}$\\ 
\hline
       &      & all & $182$ & $133$ & ${\scriptstyle\pm  5.5}$ &  $141$ & ${\scriptstyle\pm  4.4}$&  $ 59$ & ${\scriptstyle\pm  3.1}$\\ 
       & total&   V & $132$ & $158$ & ${\scriptstyle\pm  6.5}$ &  $152$ & ${\scriptstyle\pm  5.2}$&  $ 60$ & ${\scriptstyle\pm  3.7}$\\ 
       &      &  IV & $ 36$ & $107$ & ${\scriptstyle\pm 10.5}$ &  $107$ & ${\scriptstyle\pm  8.4}$&  $ 50$ & ${\scriptstyle\pm  5.9}$\\ 
\cline{2-10}
A4--A6 &      & all & $137$ & $138$ & ${\scriptstyle\pm  6.0}$ &  $144$ & ${\scriptstyle\pm  4.8}$&  $ 56$ & ${\scriptstyle\pm  3.4}$\\ 
       &normal&   V & $ 97$ & $159$ & ${\scriptstyle\pm  7.2}$ &  $157$ & ${\scriptstyle\pm  5.7}$&  $ 56$ & ${\scriptstyle\pm  4.0}$\\ 
       &      &  IV & $ 27$ & $107$ & ${\scriptstyle\pm 11.3}$ &  $110$ & ${\scriptstyle\pm  9.0}$&  $ 47$ & ${\scriptstyle\pm  6.4}$\\ 
\cline{2-10}
       & CP   &     & $ 19$ & $107$ & ${\scriptstyle\pm 19.8}$ &  $106$ & ${\scriptstyle\pm 15.8}$&  $ 69$ & ${\scriptstyle\pm 11.2}$\\ 
       & CB   &     & $ 26$ & $141$ & ${\scriptstyle\pm 15.4}$ &  $147$ & ${\scriptstyle\pm 12.3}$&  $ 63$ & ${\scriptstyle\pm  8.7}$\\ 
\hline
       &      & all & $190$ & $134$ & ${\scriptstyle\pm  5.5}$ &  $135$ & ${\scriptstyle\pm  4.4}$&  $ 60$ & ${\scriptstyle\pm  3.1}$\\ 
       & total&   V & $123$ & $149$ & ${\scriptstyle\pm  6.5}$ &  $144$ & ${\scriptstyle\pm  5.2}$&  $ 57$ & ${\scriptstyle\pm  3.7}$\\ 
       &      &  IV & $ 46$ & $110$ & ${\scriptstyle\pm 11.1}$ &  $112$ & ${\scriptstyle\pm  8.9}$&  $ 60$ & ${\scriptstyle\pm  6.3}$\\ 
\cline{2-10}
A7--A9 &      & all & $141$ & $133$ & ${\scriptstyle\pm  6.3}$ &  $134$ & ${\scriptstyle\pm  5.0}$&  $ 60$ & ${\scriptstyle\pm  3.6}$\\ 
       &normal&   V & $ 90$ & $149$ & ${\scriptstyle\pm  7.6}$ &  $143$ & ${\scriptstyle\pm  6.1}$&  $ 58$ & ${\scriptstyle\pm  4.3}$\\ 
       &      &  IV & $ 32$ & $110$ & ${\scriptstyle\pm 13.3}$ &  $114$ & ${\scriptstyle\pm 10.6}$&  $ 60$ & ${\scriptstyle\pm  7.5}$\\ 
\cline{2-10}
       & CP   &     & $ 16$ & $104$ & ${\scriptstyle\pm 23.9}$ &  $113$ & ${\scriptstyle\pm 19.1}$&  $ 76$ & ${\scriptstyle\pm 13.5}$\\ 
       & CB   &     & $ 33$ & $149$ & ${\scriptstyle\pm 10.7}$ &  $149$ & ${\scriptstyle\pm  8.5}$&  $ 49$ & ${\scriptstyle\pm  6.0}$\\ 
\hline
       &      & all & $197$ & $107$ & ${\scriptstyle\pm  5.2}$ &  $113$ & ${\scriptstyle\pm  4.2}$&  $ 59$ & ${\scriptstyle\pm  3.0}$\\ 
       & total&   V & $125$ & $117$ & ${\scriptstyle\pm  6.7}$ &  $122$ & ${\scriptstyle\pm  5.3}$&  $ 59$ & ${\scriptstyle\pm  3.8}$\\ 
       &      &  IV & $ 55$ & $ 99$ & ${\scriptstyle\pm  9.7}$ &  $102$ & ${\scriptstyle\pm  7.7}$&  $ 57$ & ${\scriptstyle\pm  5.5}$\\ 
\cline{2-10}
F0--F2 &      & all & $150$ & $110$ & ${\scriptstyle\pm  5.9}$ &  $114$ & ${\scriptstyle\pm  4.7}$&  $ 57$ & ${\scriptstyle\pm  3.3}$\\ 
       &normal&   V & $ 91$ & $117$ & ${\scriptstyle\pm  7.4}$ &  $123$ & ${\scriptstyle\pm  5.9}$&  $ 57$ & ${\scriptstyle\pm  4.2}$\\ 
       &      &  IV & $ 44$ & $100$ & ${\scriptstyle\pm 11.3}$ &  $108$ & ${\scriptstyle\pm  9.0}$&  $ 60$ & ${\scriptstyle\pm  6.4}$\\ 
\cline{2-10}
       & CP   &     & $ 18$ & $ 84$ & ${\scriptstyle\pm 18.7}$ &  $ 92$ & ${\scriptstyle\pm 14.9}$&  $ 63$ & ${\scriptstyle\pm 10.5}$\\ 
       & CB   &     & $ 29$ & $123$ & ${\scriptstyle\pm 14.3}$ &  $120$ & ${\scriptstyle\pm 11.4}$&  $ 61$ & ${\scriptstyle\pm  8.1}$\\ 
\hline
\end{tabular}
  \end{center}
\end{table}

\begin{table}[!htbp]
  \begin{center}
\caption{{\bf (extract)}
List of the 1541 B9- to F2-type stars, with their \vsini\ value, spectral type, associated subgroup and classification (CP, CB, blank stands for ``normal'').}
\label{list} 
\setlength\tabcolsep{4pt}
\begin{tabular}{rrlll}
\hline
\hline
\multicolumn{1}{c}{HD}  & \multicolumn{1}{c}{\vsini}   & \multicolumn{1}{c}{Sp. type} & \multicolumn{1}{c}{subgroup} & \multicolumn{1}{c}{classification} \\
  & \multicolumn{1}{c}{(\kms)}  \\
\hline
     3 & 228 & A1Vn   &      A0--A1 &  \\
   203 & 170 & F2IV   &      F0--F2 &  \\
   256 & 241 & A2IV/V &      A2--A3 &  \\
   319 &  59 & A1V    &      A0--A1 & CP\\
   431 &  97 & A7IV   &      A7--A9 & CB\\
   $\cdots$\\
\hline
\end{tabular}
  \end{center}
\end{table}

Several trends are noticeable in the \vsini\ distributions, as seen in Fig.~\ref{distrib1}:
\begin{itemize}
\item[$\bullet$] The \vsini\ distributions for most massive stars (B9 and A0--A1) exhibit hints of a bimodality, even when only taking into account ``normal'' stars.
A double peak in the $v\sin i$ of A0V-type stars was pointed out by \citet{Raa_89}. The distributions for B9 and A0--A1 groups are very similar. This similarity becomes quite clear if the histograms normalized to the number of stars in each group are compared (as they are displayed in Fig.~\ref{distrib2}). There are two maxima, one near $\vsini \simeq 160 \,\kms$ and the other near $v\sin i \simeq 50 \,\kms$. The excess of stars with small \vsini\ cannot be considered due to Am and Ap stars, as most of them were taken off from the sample.

\item[$\bullet$] There is a regular decrease of the maximal values of \vsini\ from B9 to F0--F2 groups, which is only marginally correlated with the critical equatorial velocity $v_\mathrm{c} \propto (\mathscr{M}/R)^{1/2}$; i.e. $v_\mathrm{c}(\mbox{B9})/v_\mathrm{c}(\mbox{F0--F2}) \sim 1.1$, while $\vsini_\mathrm{max}(\mbox{B9})/\vsini_\mathrm{max}(\mbox{F0--F2}) \sim 1.4$.

\item[$\bullet$] The proportion of low \vsini\ stars ($\la 70$\,\kms), in the full subsamples, diminishes from B9 to A6-type stars (B9: 31\,\%; A0--A1:  31\,\%; A2--A3: 24\,\%; A4--A6: 10\,\%) and then increases towards the later types (A7--A9:  14\,\%; F0--F2: 24\,\%).  There is a striking lack of stars with $\vsini <50\,\kms$ in the A4--A6 group. The same tendency is found, considering ``normal'' stars, but the proportion is lower.
 The effect of the proportion of low \vsini\ stars is also reflected in the variation of the dispersion (see Table~\ref{moments}). It decreases from B9 to A4 stars, and reaches a plateau around 60\,\kms\ as soon as the distribution is not significantly skewed.

\end{itemize}

From this point, the assumption that stellar rotation axes are randomly oriented is adopted. This hypothesis has been tested many times \citep{Gry92,Gae93} and is still the most valid.
{\bf The following sections focus on the ``normal'' stars, excluding the stars classified as CP or CB in the previous section.}


\section{Distributions of equatorial rotational velocities}
\label{Sect_vdist}
\subsection{Rectified distributions}
\label{rectification}
The observed \vsini\ parameter is the projection of the equatorial velocity $v$ of the star on the line of sight, $i$ being the inclination between the stellar rotation axis and the line of sight.
The Probability Density Function (hereafter PDF) of the \vsini\ is thus the result of the convolution between the distribution of ``true'' equatorial velocities $v$, the distribution of inclination angles $i$, and the observational error law.

Let $\vartheta$ be the true projected rotational velocity of a star, the PDF of \vsini\ can be written as:
\begin{equation}
\label{eq:conv1}
\Phi(\vsini) = \int \Psi(\vartheta)\,P(\vsini|\vartheta)\, \mathrm{d}\vartheta,
\end{equation} where $P(\vsini|\vartheta)\,\mathrm{d}\vartheta$ is the conditional probability of $\vsini\in[\vartheta,\vartheta+\mathrm{d}\vartheta]$. $\Psi(\vartheta)$ is the PDF of ``true'' projected velocities, which itself can be written as:
\begin{equation}
\label{eq:conv2}
\Psi(\vartheta) = \int \Upsilon(v)\,P(\vartheta|v)\, \mathrm{d}v,
\end{equation} where $P(\vartheta|v)\,\mathrm{d}v$ is the conditional probability of $\vartheta\in[v,v+\mathrm{d}v]$, and $\Upsilon(v)$ is the PDF of the ``true'' equatorial velocities.

The aim of this section is to recover the distribution $\Upsilon(v)$ from the observations.

\subsubsection{Smoothing of the distributions}
The observed PDF $\Phi(\vsini)$ can be estimated from the \vsini\ data. For that purpose, the kernel estimator $\hat{\Phi}$, with kernel $K$, is used. Details can be found in Appendix~\ref{Notes2}. The estimator $\hat{\Phi}$ is defined as:
\begin{equation}
\label{kernelestimator}
  \hat{\Phi}(x) = {1\over n}\sum_{i=1}^{n}{1\over h}\, K\left({x -X_i \over h}\right),
\end{equation} where $n$ is the sample size and $h$ is the window width, also called the
smoothing parameter.
The error on \vsini\ is expected to be multiplicative, as shown in Papers~I and II. This can be grasped by considering the \vsini\ as a scale/dilatation parameter, like the dispersion (whereas the radial velocity is a shift parameter like the mean).The error on the empirical dispersion is asymptotically proportional to the dispersion (see Appendix~\ref{Notes0}).
The natural error law would thus be a log-normal distribution for the \vsini. In Paper~I, the distribution of observed \vsini\ for Sirius is perfectly fitted either by a normal or a log-normal distribution.
This noticing triggers the processing of this step in logarithmic space of velocities, to get rid of the variance heterogeneity \citep{Zar99}. The data are significantly higher than zero to safely use  $\ln(\vsini)$ instead of $\ln(\vsini + 1)$ \citep{Bat47}.
The smoothing of the PDF is then performed on logarithmic velocities, and the PDF $\Phi_\mathrm{log}( \ln(\vsini))$ is computed by the Kernel method, using a Gaussian kernel (standard normal distribution):
\begin{equation}
\label{gaussiankernel}
 K(t) =  {1\over\sqrt{2\,\pi}}\,\mathrm{e}^{-{1\over 2}{t^2}}.
\end{equation}
 The smoothing parameter $h$ in Eq.~\ref{kernelestimator} is estimated using the scheme described by \citet{ShrJos91}, and based on each subsample data. The estimator $\hat{h}$, as solution of the Eq.~12 in \citet{ShrJos91}, is given in Table~\ref{sjeqd} for each subsample.

\begin{table}[!htbp]
\begin{center}
\caption{Estimated bandwidth for the kernel method and mean integrated squared error: for each subsample, the size $n$ (number of stars) is given, as well as the estimated $\hat{h}$ according to \citet{ShrJos91}, computed in the logarithmic velocity scale, and the variability band width $\epsilon$ (Eq.~\ref{epsilon}).}
\label{sjeqd}
\begin{tabular}{lrrr}
\hline
\hline
\multicolumn{1}{c}{Subsample} & \multicolumn{1}{c}{$n$} & \multicolumn{1}{c}{$\hat{h}$}  & \multicolumn{1}{c}{$\epsilon$} \\
\hline
B9     & $125$ & $0.223$ & $0.0503$ \\
A0--A1 & $271$ & $0.147$ & $0.0421$ \\
A2--A3 & $258$ & $0.170$ & $0.0401$ \\
A4--A6 & $137$ & $0.152$ & $0.0583$ \\
A7--A9 & $141$ & $0.137$ & $0.0604$ \\
F0--F2 & $150$ & $0.149$ & $0.0561$ \\
\hline
\end{tabular}
\end{center}
\end{table}

It must be emphasized, however, that using a unique value $\hat{h}$ for a given sample of logarithmic \vsini\ data will lead to create artificial and non significant distribution modes at low \vsini. For low \vsini\ values, the bandwidth $h$ constant in logarithmic space is underestimated, due to the limitation of the precision on derived \vsini. In addition, for some of the subsamples, the number of slow rotators is small, which contributes to the presence of non significant peaks in the distributions. These peaks are clearly visible in the linear velocity space (Fig.~\ref{distrib2}).

\begin{figure}[!htp]
\resizebox{\hsize}{!}{\includegraphics{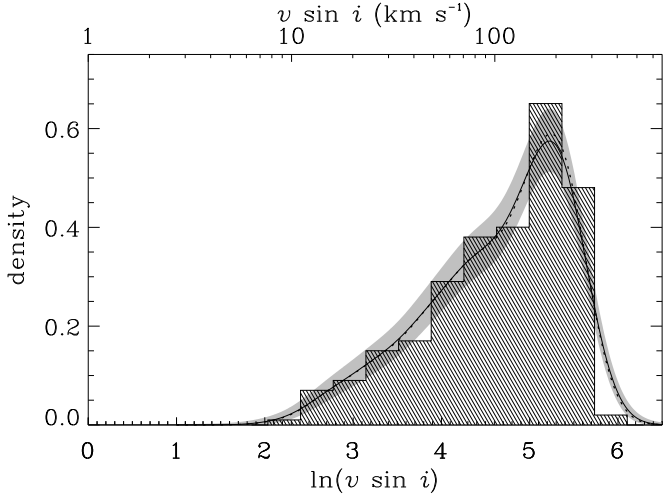}}
\hfill
\caption{Distributions of the logarithmic rotational velocities $\ln(\vsini)$ for the A0--A1 subsample. The corresponding scale of \vsini\ is plotted as the upper x-axis. The histogram represents the data for ``normal'' stars, displayed in Fig.~\ref{distrib1}. The solid line is the smoothed distribution $\Phi_\mathrm{log}(\ln(\vsini))$, obtained by using a Kernel method with a Gaussian kernel, on the $\ln(\vsini)$. The thick dotted line stands for the distribution $\Psi_\mathrm{log}(\ln\vartheta)$ of the ``true'' logarithmic velocities, rectified from a Gaussian error law (see text).}
\label{rectif1}
\end{figure}

\citet[ Chap 2.3]{BonAzi97} propose a way of quantifying the variability of the density estimates and therefore assessing the significance of the present modes. Using this method, only variance is assessed to compute the resulting ``variability bands'', but the graphical display of this variance structure allows the disentanglement between the genuine features and the ``noisy'' ones.
The variability bands are estimated by the $\epsilon$ parameter (Table~\ref{sjeqd}, see Appendix~\ref{Notes2}) and are over-plotted as gray areas in Figs~\ref{rectif1} and \ref{distrib2}.

The PDFs $\Phi_\mathrm{log}(\ln(\vsini))$, for each subsample, are the result of the kernel density estimations. The result for the largest subsample (A0--A1) is shown in Fig.~\ref{rectif1}.

\subsubsection{Rectification of the error law effect}

This operation is an intermediate step in order to recover the PDF of true equatorial velocities.
The rotational velocities and its associated error can be written as:
\begin{equation}
\label{relerror}
\vsini \pm \alpha\,\vsini \;\approx\; \vsini\,\td\,
  (1+\alpha),
\end{equation} for small values of $\alpha$ (i.e. terms of second order are negligible), the error can be written on a multiplicative form\footnote{\td\ is the symbol times/divide, multiplicative equivalent of plus/minus $\pm$}. Therefore the logarithm scale can be considered: $\ln(\vsini) \pm \ln(1+\alpha)$.
In this case, the coefficient $\alpha$ is taken to be 0.1. In Papers~I and II, it was respectively 0.06 and 0.05 for \vsini\ derived from ECHELEC and AUR\'ELIE data, but for the \vsini\ measured by Abt, the relative error is closer to 10\,\%. Using  $\alpha=0.1$, the second order is obviously negligible, and a 1\,\%-effect on the observed \vsini\ is really not significant.

The error of $\ln(\vsini)$ is assumed to be normally distributed with a mean of $0$ and a variance of ${\left(\ln(1+\alpha)\right)}^2$. The conditional probability $P(\ln(\vsini)|\ln\vartheta)$ takes the shape of a Gaussian function:
\begin{equation}
\label{probconde}
  P(\ln(\vsini)|\ln\vartheta) = {1\over\sqrt{2\,\pi}\,\ln(1+\alpha)}
  \mathrm{e}^{-{1\over 2}{\left(\ln(\vsini)-\ln\vartheta\over\ln(1+\alpha)\right)}^2},
\end{equation}
in the equation:
\begin{equation}
 \Phi_\mathrm{log}(\ln(\vsini)) = \int \Psi_\mathrm{log}(\ln\vartheta)\,P(\ln(\vsini)|\ln\vartheta)\, \mathrm{d}\ln\vartheta.
\end{equation}

$\Psi_\mathrm{log}(\ln\vartheta)$ is derived using the \citet{Luy74} iterative technique. Further details are given in Appendix~\ref{Notes3}. The PDF $\Phi_\mathrm{log}( \ln(\vsini))$ and $\Psi_\mathrm{log}(\ln\vartheta)$ of the subsample A0--A1 are shown in Fig.~\ref{rectif1} and illustrate the steepening and sharpening of the distribution deconvolved by the error law.

The resulting distributions are transformed back in the linear velocity space following:
\begin{equation}
  \Psi(\vartheta) = {\Psi_\mathrm{log}(\ln\vartheta)\over\vartheta}.
\end{equation}

\begin{figure*}[!htp]
\resizebox{\hsize}{!}{\includegraphics{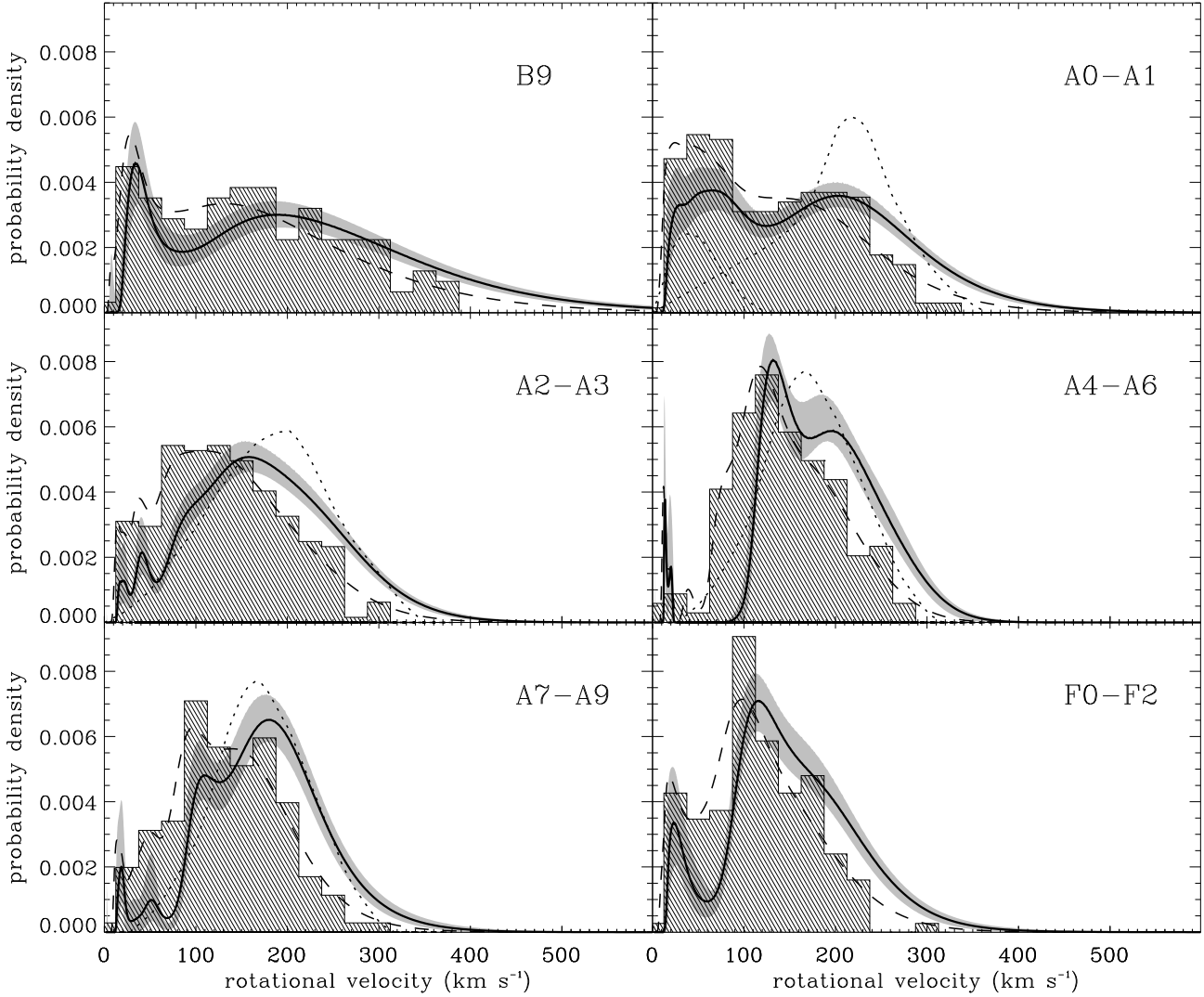}}
\hfill
\caption{Distributions of rotational velocities for the observed subsamples: shaded histograms are the observed \vsini\ and are identical to the histograms for ``normal'' stars in Fig.~\ref{distrib1}; dashed lines stand for the smoothed distributions of true projected rotational velocities $\Psi(\vartheta)$; solid thick lines are the distributions of true equatorial velocities $\Upsilon(v)$ and the gray strips are their associated variability bands.
Dotted lines are the distributions found by AM: in our A0--A1 sample both peculiar and normal star distributions are overplotted, in our A2--A3 sample the distribution of A3--A4 stars found by \citet{AbtMol95} is overplotted, and in our A4--A6 and A7--A9 samples the distribution of A5--F0 stars is shown.}
\label{distrib2}
\end{figure*}

The distributions of ``true'' projected velocities $\Psi(\vartheta)$ are displayed for all the subsamples in Fig.~\ref{distrib2}.

\subsubsection{Rectification of the projection effect}

Under the assumption of randomly oriented rotation axes, the
conditional probability $P(\vartheta|v)$ in Eq~\ref{eq:conv2} is:
\begin{equation}
\label{probcondi}
  P(\vartheta|v) = {\vartheta\over v}\,{H(v-\vartheta)\over\sqrt{v^2-\vartheta^2}},
\end{equation} where $H$ is the Heaviside function\footnote{The Heaviside step
  function $H$ is defined by:
\begin{eqnarray*}H(x) = \left\{\begin{array}{ccc}
1& \mathrm{if} & x\geq0\\
0&  \mathrm{if} & x< 0
\end{array}\right.\end{eqnarray*}}.

Combining Eqs~\ref{eq:conv2} and \ref{probcondi}, the PDF of $\vartheta$
becomes:
\begin{equation}
\label{abeleq}
 \Psi(\vartheta) = \int_\vartheta^\infty \Upsilon(v)\,{\vartheta\over v\,\sqrt{v^2-\vartheta^2}}\, \mathrm{d}v,
\end{equation} which is an Abelian integral that can be analytically solved.
Details about this Abelian form of the convolution can
be found in \citet{ChrMuh50}. \citet{Jon_01} also describe it in the
purpose of the deconvolution of the distribution of projected masses.

Therefore two ways are possible to compute the final distribution of
equatorial velocities $\Upsilon(v)$.
In this work, only the Lucy-iteration technique \citep{Luy74} has been used.

These distributions $\Upsilon(v)$ are displayed in Fig.~\ref{distrib2} for the six subsamples, and described in the next Section.

\subsection{Description of the $v$ distributions}

The distributions of true equatorial velocities imply the following facts:
\begin{itemize}

\item[$\bullet$] The bimodal character of distributions corresponding to B9 and A0--A1 field dwarf stars is clearly present, even the samples have been cleared from all known Am, Ap and close binaries.
For B9-type stars: the mode of slow rotators is $\sim 35$\,\kms,  and for fast rotators $\sim 190$\,\kms.
For the A0--A1 subsample: they are $\sim 60$\,\kms\ and $\sim 200$\,\kms\ respectively.
The modes of fast rotators ($v\ga 150$\,\kms), for B9 and A0--A1 type stars, are fitted by Maxwellian PDFs (Fig.~\ref{maxwell}).
As shown by \citet{Deh67}, the distributions of $v$ are well fitted by Maxwellian functions:
\begin{equation}
\Upsilon_\mathrm{fast}(v) = {\sqrt{2/\pi}\over a^{3/2}}\,v^2\,\mathrm{e}^{-v^2\over 2a}.
\end{equation}
The parameter $a$ is defined by the mode of the distribution $\sqrt{2a}$.
The proportion of slow rotators ($v\la 150$\,\kms) is taken as the excess compared to the Maxwellian fast rotator distribution. The results are displayed in Table~\ref{tabmaxwell}. The slow rotator peak corresponds to some 19 stars for B9-type stars, and about 66 stars for A0--A1-type stars.

\begin{table}[htbp]
  \begin{center}
\caption{Characteristics of the bimodal distributions: the fast rotator mode $\sqrt{2a}$ from the Maxwellian distribution parameter $a$, the proportion $p$ of slow rotators as the excess compared to the Maxwellian fast rotator distribution, the corresponding number $N_\mathrm{s}$ of slow rotators in the observed sample, and the mean of the slow rotator distribution $\mu_\mathrm{s}$.}
\label{tabmaxwell}
    \begin{tabular}{lccccc}
\hline
\hline
Spec. types & $\sqrt{2\,a}$ &  $p$ & $N_\mathrm{s}$ & $\mu_\mathrm{s}$\\
            & (\kms)        &      &                & (\kms) \\
\hline
       B9     &  $190$ & $0.15$ &  $19$ &   $46$\\
       A0--A1 &  $200$ & $0.24$ &  $66$ &   $59$\\
\hline
    \end{tabular} 
  \end{center}
\end{table}
\begin{figure}[!htbp]
\resizebox{\hsize}{!}{\includegraphics{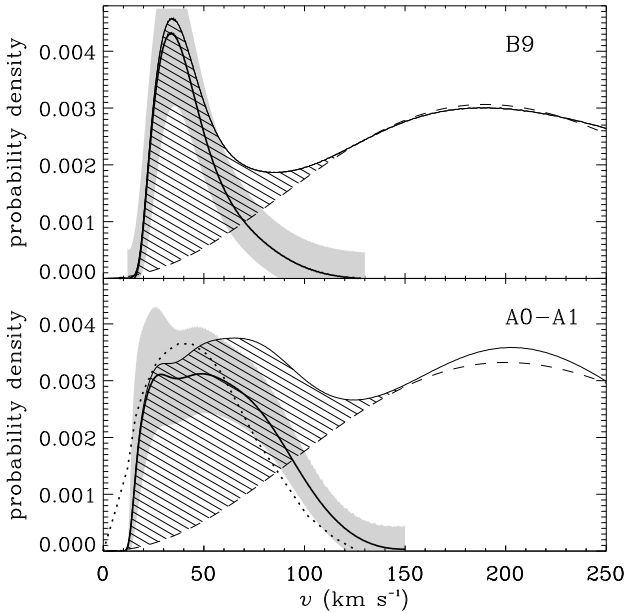}}
\caption{Slow rotators in B9 and A0--A1 type stars: the PDF of equatorial velocities $v$ (solid line) are displayed. The fast rotator mode is fitted by a Maxwellian distribution (dashed line), and the slow rotator distribution is estimated as the excess compared to the Maxwellian distribution (hatched area). The difference between both distributions is plotted (bold solid line) 
together with its propagated variability band (gray strip). For the A0--A1 type stars, the distribution of chemically peculiar objects found by AM is overplotted (dotted line), normalized to represent the same number of stars.}
\label{maxwell}
\end{figure}

\item[$\bullet$] In the A2 to A9 spectral type groups, the small wiggles of distributions in the velocity interval $0 \la v \la 70$\,\kms\ concern a negligible fraction of stars. Moreover the variability bands associated to the distributions argue in the sense that the presence of these slow rotators is not significant. These objects are probably unknown synchronized binaries or chemically peculiar stars that pollute the sample of ``normal'' stars. There is a net lack of rotators with $v \la 70$\,\kms. An absolute minimum of this fraction is seen in the A4--A6 group.

\begin{figure}[!htp]
\resizebox{\hsize}{!}{\includegraphics{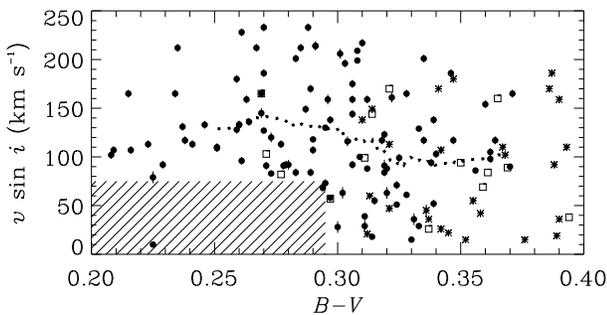}}
\caption{
Projected rotational velocities as a function of the $B-V$ index (From HIPPARCOS catalog) of the 150 ``normal'' stars in the F0--F2 subgroup. Filled circles ($\bullet$) stand for F0-type stars, open squares ($\Box$) for F1 and asterisks (*) for F2-type stars. The hatched strip emphasizes the lack of object in the area ($\vsini<75$\,\kms, $B-V<0.295$\,mag), the only object in this area is \object{HD~36876}. The dotted line is the 40-point running average of \vsini\ for the 150 objects.}
\label{F0F2}
\end{figure}

\item[$\bullet$]  An excess of slow rotators with velocities $0 \la v \la 60$\,\kms\ appears in the F0--F2 subgroup. This mode of slow rotators, around $20$\,\kms, in the distribution for early-F stars, is significant.
The $B-V$ distribution for all the ``normal'' F0--F2 stars, ranging from $0.2$ to $0.4$\,mag (Fig.~\ref{F0F2}), shows that all (but one) stars with $\vsini<75$\,\kms\ have color index $B-V\ga 0.3$\,mag. It ensues that the slow rotator peak ($0 \la v \la 60$\,\kms) is only composed of stars redder than $B-V= 0.3$\,mag.
 The drop of rotation along the main sequence through the early F-type stars was pointed out by \citet{Win66} and \citet{Krt67}. The decrease in rotational velocity observed in Fig.~\ref{F0F2} is the beginning of this rotational break. A similar decrease occurs also for class III giants \citep{Gry89}.

\item[$\bullet$] {\bf In the high-velocity side of distributions (Fig.~\ref{distrib2}) a well defined peak can be noticed at $v \simeq 200$\,\kms\ in all groups, except for the F0--F2 for which at $v \simeq 200$\,\kms\ there is only an inflection.}

\item[$\bullet$] A Roche representation of the rotating stellar surfaces is used to transform the A2 to F2 velocity distributions into $\Omega/\Omega_\mathrm{c}$-probability distributions (see Appendix~\ref{Notes5}). They are shown in Fig.~\ref{omegaprob}. It can be seen that in each group, more than half of the stars rotate with $\Omega/\Omega_\mathrm{c} \ga 0.6$ and that the marked lacks of stars are for $\Omega/\Omega_\mathrm{c} \la 0.3$. Objects with $\Omega/\Omega_\mathrm{c} \ga 0.6$ can undergo geometrical deformations induced by the rotation, which implies that their spectral characteristics may be somewhat aspect angle dependent.
\end{itemize}

\subsubsection{Comparison with AM}

Figure~\ref{distrib2} compares the distributions of true rotational velocities from AM (dotted lines) with those obtained in this work (solid lines). For the later A-type subgroups (A2--A3, A4--A6, A7--A9), there is good agreement between the distribution of normal stars in AM with ours. It suggests that our selection of non CP stars is efficient. As regards our distributions for A4--A6 and A7--A9 subgroup, compared to the corresponding A5--F0 in AM, the tails towards lower $v$ is much steeper in our study.

On the other hand a strong discrepancy is observed in the A0--A1 subgroup. The fast rotator mode found in the present work corresponds to the distribution found by AM for normal stars (Fig.~\ref{distrib2}). The excess of slow rotators emphasized in Fig.~\ref{maxwell} has a distribution similar to the Ap+Am stars found by AM, even though to discard the CP and CB stars, the same selection criteria as for later type stars were applied.

\citet{Abt00} argues that rotational velocity alone discriminates the normal A-type stars from the peculiar ones. However \citet{Mas04} points out that Abt's definition of peculiar stars encompasses stars that do not exhibit the abundances anomalies of ``classical'' CP stars.


\section{Discussion} 
\label{Sect_Disc}
\begin{figure*}[!htp]
\resizebox{\hsize}{!}{\includegraphics{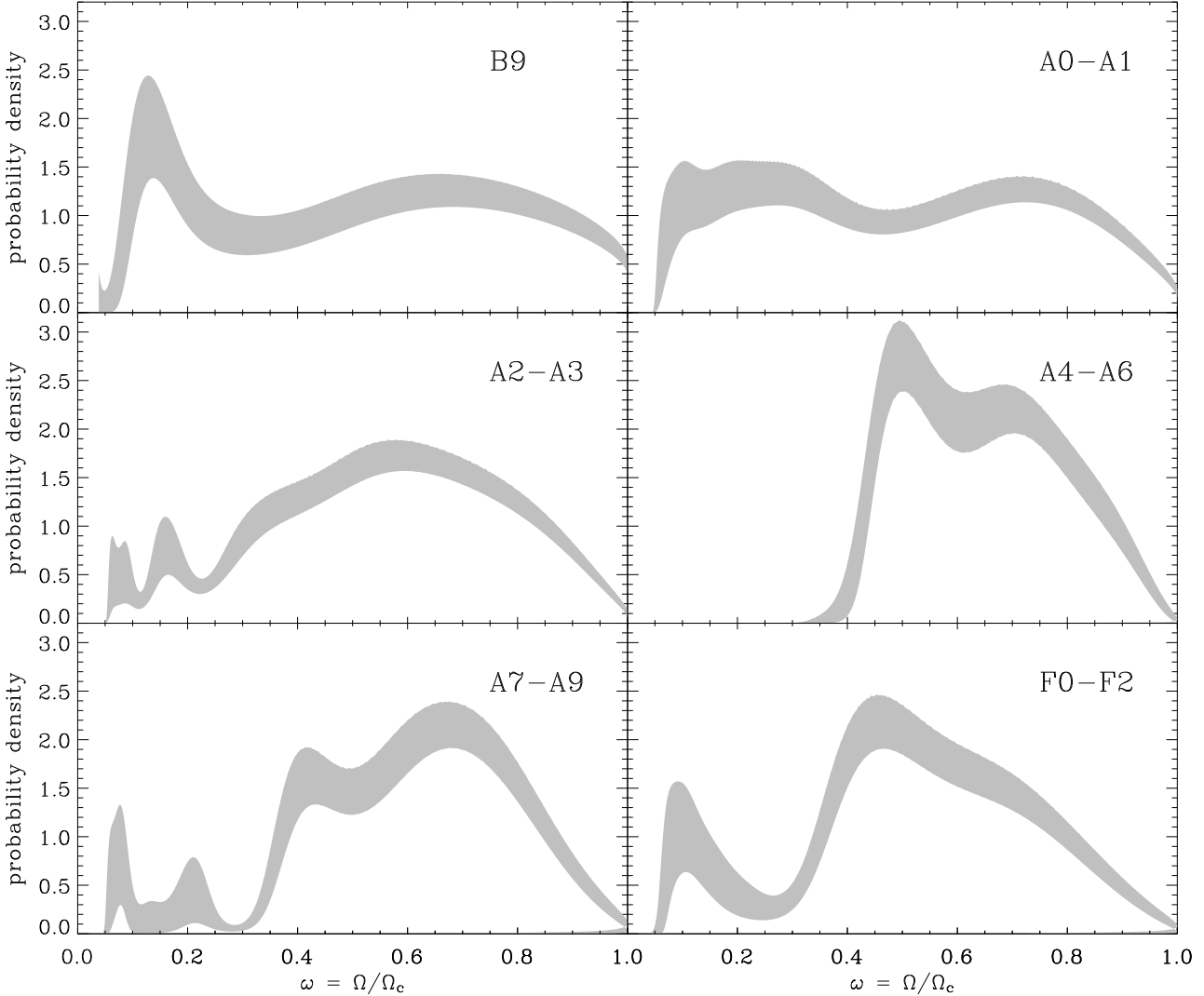}}
\caption{Distributions of angular velocities $\Omega/\Omega_\mathrm{c}$ derived from $v$ distributions in Fig.~\ref{distrib2} and using Appendix~\ref{Notes5}. Only the variability band of each PDF is displayed.}
\label{omegaprob}
\end{figure*}

The main objective of this paper was to obtain distributions of surface rotational velocities of a sample of \emph{field} stars in a spectral type range where objects with the earlier spectral types have radiative envelopes and those with cooler spectral types have convective envelopes. The present research is intended to detect signatures of multimodal rotational velocity distributions among ``normal'' dwarf stars, which are in the early MS evolutionary phases.

 Up to now, for stars with masses $1.3\!\la\!\mathscr{M}/\mathscr{M}_{\odot}\!\la\!3.0$, there are indications for bimodality only for cluster late B-type stars \citep{Gue82} and for near solar mass stars \citep[references in][]{Bas03}. In the mass range concerned by the A0--F0 stellar sample of \citet{AbtMol95}, the detected excess of low rotators was assigned entirely to Am and Ap stars. The Ap phenomenon is, however, evolutionary related, because it appears after the stars completed at least the first third of their MS life \citep{Hug_00a,Stn00}. On the other hand, most Am stars do not have luminosity class determination, so that their evolutionary status is unclear. Moreover, a great fraction of them, if not all of them, are binaries \citep{Dei00}.
The Am+Ap group cannot then be considered witnessing a neat low velocity component in a possible bimodal distribution. We cleansed our stellar sample from objects with known Am and Ap signatures, and from all known close binaries to avoid strong spurious effects on the velocity distributions carried by braking effects that can not be considered forming part of the initial MS conditions for single stars. Thus, as binaries and stars with low rotation possibly due to evolutionary effects were removed, the distributions we found for field B9, A0--A1 and F0--F2 stars may be considered as genuinely bimodal.

 The simplest explanation that can be tempted to these bimodality is that the braking of surface rotational velocities trough interactions with the circumstellar disc prior to the ZAMS phase is more or less lasting. The two velocity peaks would then imply the existence of two preferential interaction time scales, which is difficult to prove. However, this explanation sounds not sufficient, because in stars with spectral types from A2 to A9 the bimodal aspect of velocity distribution is not obvious. Perhaps, also some phenomena of internal angular momentum redistribution can be present.

 It is worth noting the lack of stars with $\Omega/\Omega_{\rm c}\!\la\!0.4$ in the A4--A6 spectral-type group. This group is roughly in the middle of the  transition from stars with radiative envelopes to those with convective envelopes. The behavior of the $1600\,$\AA\ intensity jump \citep{Boe82} in stars from A1 to A7, reveals a strong ionization balance change in the envelope, and so in their structural patterns, as we pass from one spectral type to another. It can thus happen that either the PMS angular momentum setting-up phenomena are somehow related with stellar structure and its external ionization balance, or that in the early ZAMS phases some internal redistribution of angular momentum operates making that $\Omega/\Omega_{\rm c}\!\la\!0.4$ {\bf is unlikely}.

 On the other hand, in all subgroups there is a rough average of $15\,\%$ stars with $\Omega/\Omega_{\rm c}\!\ga\!0.8$. These stars are subjected to rotationally induced geometrical distortions and to gravitational darkening. Since the local effective temperature is, according to the von Zeipel theorem, $T_{\rm eff}^4\!\propto\!g$ ($g$ is the local stellar surface gravity) in stars with radiative envelopes and $T_{\rm eff}^4\!\propto\!g^{0.3}$ \citep{Luy67,Pez_99} in objects with convective envelopes, a strong mixing of stars with different masses is expected in the hottest stellar groups. The gravitational darkening effect can also play a role in the selection of objects in the A4--A6 spectral-type group. In fact, pole-on stars rotating at rates $\Omega/\Omega_\mathrm{c}\!\sim\!0.8$ which are seen as A3 (average hemisphere effective temperature $T_{\rm eff}^{\rm pole}\!\simeq\!8700$~K), would be seen equator-on likely as A7 (average hemisphere effective temperature $T_{\rm eff}^{\rm equator}\!\simeq\!7900$~K). In non-rotating stellar models, the effective temperature range $7900 \la\!T_{\rm eff}\!\la\!8700$~K corresponds to the transition from stars with radiative envelopes to those with convective envelopes. 
It would then be interesting to investigate if such objects present a radiative/convective structural dichotomy and if the lack of low rotators in the A4--A6 group is somehow related with this structural characteristic.

    Since there is a correlation between the strength of magnetic fields, low rotation and presence of the CP character in A to B-type stars \citep{Mas04}, a thorough inquire on the nature of objects with low rotation in our sample that do not have apparent CP aspects appears timely to address. However, the separation of stars in our sample with genuine low equatorial velocity $v$ from those with small $\sin i$ is not obvious. Our planned determination of detailed fundamental parameters and identification of fast rotating stars with low $\sin i$ with model atmospheres for fast rotators may perhaps be of some help to this purpose.

\section{Summary and conclusion}
\label{Sect_Conc}

This work was based on a large and homogeneous sample of \vsini\ measurements of B9 to F2-type gathered from \citet{Ror_02b} and \citet{Abt_02}. This sample was carefully cleaned of evolved stars and known chemically peculiar and/or binary stars, resulting in a sample of about $1100$ main-sequence single stars. This stellar sample was divided into six spectral type groups, chosen so as to warrant in each statistical significance and mass resolution, to allow for detection of possible mass dependencies of velocity distributions.

The distributions of true equatorial velocities have been obtained from observed \vsini\ distributions using the iterative technique described by \citet{Luy74} to rectify from the error and random inclination distributions.

\begin{itemize}

\item The mean true equatorial linear rotational velocity of A2 to F2 MS dwarf stars are higher than quoted in the literature. The differences with the previous estimates can attain $\Delta v \ga 50 \kms$.

\item The distribution of rotational velocities of B9 to A1-type ``normal'' field dwarf stars are genuinely bimodal. This finding is contrary to the results obtained by \citet{Gue82}, \citet{AbtMol95} and \citet{Abt00}. The presence of modes around $50$ and $200$\,\kms\ may be due to formation processes and phenomena of angular momentum loss and redistribution undergone during PMS phases.

\item There is a striking lack of rotators with $v \la 50\mbox{--}70\,\kms$ from A2 to A9-type MS field stars. The minimum of stars in this interval is for A4--A6-type stars.

\item The F0--F2 group seems to have a bimodal distribution.

\item A great fraction of A-type stars are fast rotators: $\Omega/\Omega_\mathrm{c} \ga 0.5$. Hence, it should not be excluded that the faster rotating A-type stars, mainly A4--A6-types, might have envelopes that have non homogeneous structures in latitude: radiative (polar regions) and convective (equatorial regions).
\end{itemize}

Further detailed analysis of the present stellar sample using fundamental parameter determinations and models of rotating stars will be given in forthcoming papers.

\begin{acknowledgements}
We profoundly thank Dr~H.~Levato for the early communication of the B stars rotational velocities.
Discussions with Dr~F.~Arenou were once again very useful and helpful
for the statistical approach of the work, and we also thank him for giving access to the code computing the kernel estimator.
We are deeply grateful to Prof.~S.~Sheather for providing the subroutine which estimates the data-based bandwidth.
 We warmly thank the comments and suggestions by an anonymous referee that helped to improve the presentation of the paper.
\end{acknowledgements}

\appendix
\section{Notes on common {\boldmath\vsini} standard stars}
\label{Notes1}
The four common B-type stars between  measurements from Papers~I and II and the standard values from \citet{Slk_75} are listed in Table~\ref{vsini_litt}. There are few available recent data about any \vsini\ measurement for these objects. Some details are given for them below:
\begin{itemize}
\item \object{$\beta$~Tau} has been observed at high angular resolution by \citet{RiiPen02} and no suspicion of binarity has been found. The resulting uniform disk angular diameter is  $1.56\pm 0.11$\,mas.
\item \object{$\tau$~Her} was discovered by HIPPARCOS to be a slowly pulsating B star with a period of $1\fd 24970 \pm 0\fd 00008$ \citep{Hip}. \citet{MaaHia00} recently observed its line-profile variation on \ion{He}{i} 4471 and \ion{Mg}{ii} 4481\,\AA. These variations could induce some measurement effects on the \vsini\ and explain the discrepancy found between this work and the quoted results in the literature.
\item \object{$\gamma$~Lyr}  is described in Appendix of Paper~II. 
\item \object{$\omega^2$~Aqr} is described in Appendix of Paper~I.
\end{itemize}

\section{Error on the empirical dispersion}
\label{Notes0}

The empirical variance is estimated by $s^2 =
  n^{-1}\,\sum_i x_i^2,$ where $x_i$ are supposed to have a dispersion
  $\sigma$. The variance of this estimator is:
  \begin{equation}
  \mathrm{var}(s^2)=n^{-2}\,\sum_i \mathrm{var}(x_i^2).
\end{equation}
 The  dispersion of $x_i^2$ is
 \begin{eqnarray}
 \begin{array}{ccl}
             & \sigma_{x_i^2}      &\approx 2\,|x_i|\,\sigma,\\
\mathrm{then}& \mathrm{var}(x_i^2) &\approx 4\,x_i^2\,\sigma^2.
\end{array}
\end{eqnarray}
It easily  follows that
\begin{eqnarray}
\begin{array}{ccl}\mathrm{var}(s^2)& \approx &  4\,\sigma^2\,n^{-2}\,\sum_i x_i^2\\
                                                    & =       &  4\,\sigma^2\,n^{-1}\,s^2 \\
                                                    & \approx &  4\,\sigma^4\,n^{-1},
\end{array}
\end{eqnarray}
  because $\lim_{n\to\infty} s = \sigma$. Thus
  $\sigma_{s^2}=2\,\sigma^2\,n^{-1/2}.$

  The dispersion of the empirical standard deviation $s$ is therefore:
\begin{eqnarray}
\begin{array}{ccl}
\sigma_s & \approx &\sigma_{s^2}\,(2\,s)^{-1}\\
         &  =      & \sigma^2\,n^{-1/2}\,s^{-1}\\
         & \approx & \sigma\,n^{-1/2} \propto \sigma.
\end{array}
\end{eqnarray}
Q.e.d.

\section{Notes on manipulating distributions}
\subsection{Kernel method}
\label{Notes2}
This method for density estimation is extensively described in \citet{Sim86} and \citet{BonAzi97}. Applications can also be found in the literature \citep[see e.g.][]{Aru93,Jon_01}.
Using the kernel function $K$, the probability density function $\Phi$ is estimated by:
\begin{equation}
\label{kernelestimatorApp}
  \hat{\Phi}(x) = {1\over n}\sum_{i=1}^{n}{1\over h}\, K\left({x -X_i \over h}\right),
\end{equation} where $n$ is the sample size and $h$ is the band width parameter.
The mean integrated squared error (MISE) is used to measure how efficient $\hat{\Phi}$ is in estimating $\Phi$. It is defined as:
\begin{equation}
  \mathrm{MISE}(\hat{\Phi}) = E\int\left\{\hat{\Phi}(x)-\Phi(x)\right\}^2\mathrm{d}x,
\end{equation} where $E\{f(x)\}$ denotes the expectation value of a function $f(x)$ in a variable $x$. 
Minimizing the MISE allows the choice of the best kernel function. Although it is well known that the so-called {\em Epanechnikov kernel}\footnote{The Epanechnikov kernel $K_\mathrm{E}$ is defined by:
\begin{eqnarray*} K_\mathrm{E}(t)= \left\{\begin{array}{cl}
\displaystyle{3\over 4\sqrt{5}}\left(1-\displaystyle{1\over 5}t^2\right)& \mathrm{if}  -\sqrt{5}<t<\sqrt{5},\\
0&  \mathrm{otherwise.}\end{array}\right.\end{eqnarray*}} is the kernel function minimizing the MISE \citep{Sim86}, the Gaussian kernel, with a lower efficiency, has been chosen in order to make use of the \citet{ShrJos91} method for estimating the bandwidth parameter $h$. 
This latter parameter is estimated using the Fortran subroutine available from \citet{ShrJos91}. The kernel estimator $\hat{\Phi}$ is computed using the C code written by \citet{Aru93}.

The {\em variability bands}, as described by \citet[ Chap 2.3]{BonAzi97}, are computed to assess the significance of the modes in the derived distributions. For a given estimated density function $\hat\Phi$ defined by Eq.~\ref{kernelestimatorApp}, and according to \citet{BonAzi97}, the variability band width around $\sqrt{\hat\Phi}$ is:
\begin{equation}
2\,\epsilon = \sqrt{{(n\,h)}^{-1}\int K^2(t) \mathrm{d}t}.
\end{equation}
In the case of a Gaussian kernel function (Eq.~\ref{gaussiankernel}), it becomes:
\begin{eqnarray}
\label{epsilon}
2\,\epsilon &= &\sqrt{{\left(n\,h\right)}^{-1}\,\left(2\sqrt{\pi}\right)^{-1}}\nonumber\\
\mathrm{therefore}\qquad \epsilon &\approx & {0.2656\over\sqrt{\hat{h}\,n}}.
\end{eqnarray}

This is applied to the estimated PDFs $\hat{\Phi}_\mathrm{log}(\ln(\vsini))$ in Sect.~\ref{rectification}, and the corresponding $\epsilon$ values are given in Table~\ref{sjeqd} for each estimated $\sqrt{\hat{\Phi}_\mathrm{log}}$.
The variability band of $\hat{\Phi}_\mathrm{log}$ is bounded by both functions ${\left(\sqrt{\hat{\Phi}_\mathrm{log}}\pm\epsilon\right)}^2$.

These bands, first computed for the smoothed estimated distributions, are propagated through the whole processing (Lucy-iteration technique) from smoothed density functions to final deconvolved ones.
\subsection{Lucy iterative technique}
\label{Notes3}

\citet{Luy74} gives an iterative deconvolution scheme to solve the problems that consist in estimating the density function $g(Y)$ of a
quantity $Y$ from a sampled population characterized by:
\begin{equation}
    \label{eq:conv3}
 f(x) = \int g(y)\,P_X(x|y)\, \mathrm{d}y,
\end{equation}

The first guess of $g(y)$, $g^0(y)$, is taken to be the uniform law.
According to Lucy, the iterated $r$-th functions are:
\begin{equation}
    \label{eq:f_r}
 f^r(x) = \int g^r(y)\,P_X(x|y)\, \mathrm{d}y,
\end{equation}

\begin{equation}
    \label{eq:g_r+1}
 g^{r+1}(y) = g^r(y)\,\int {\hat{f}(x)\over f^r(x)}\,P_X(x|y)\, \mathrm{d}x,
\end{equation}
where $\hat{f}(x)$ is the estimation of the PDF $f(x)$ based on the observed data.

These operations are carried on using the smoothed distributions and overlapping parabolas integration method \citep{DasRaz84}.

To stop the iterations, Lucy used a $\chi^2$ test. We preferred to use a Kolmogorov-Smirnov test to check whether the hypothesis $f^r\equiv\hat{f}$ is true within a 1\,\%-threshold.
In the calculations carried out on the \vsini\ data and detailed in Sect.~\ref{rectification}, the number of iterations of the Lucy technique is $3$ or $4$ for the deconvolution by the errors distribution $\left[f(x)=\Phi_\mathrm{log}(\ln(\vsini)), g(y)=\Psi_\mathrm{log}(\ln(\vartheta))\right]$, and ranges from $12$ to $40$ for deconvolution by the inclinations distribution [$f(x)=\Psi(\vartheta)$, $g(y)=\Upsilon(v)$].

\section{Angular velocities $\Omega/\Omega_\mathrm{c}$}
\label{Notes5}

In this representation, the variation of the polar radius with $\omega=\Omega/\Omega_\mathrm{c}$ is neglected.

The Roche model of surface equipotential for a rotating star gives:
\begin{equation}
{R_\mathrm{e}(\omega) \over R_\mathrm{c}} = 1 + 0.5\,\varepsilon,
\end{equation}
where $R_\mathrm{c}$  is the critical radius: $R_\mathrm{c} = R_\mathrm{e}(\omega=1)$, and
\begin{equation}
\varepsilon=\omega^2\,{\left({R_\mathrm{e}(\omega)\over R_\mathrm{c}}\right)}^3.
\end{equation}

As $w= \omega\,{R_\mathrm{e}(\omega)\over R_\mathrm{c}}$, it follows that:

\begin{equation}
\omega = w\,{\left(1 - 0.5\,w^2\right)}.
\end{equation}


proxima.obspm.fr:/home/royer/Science/Vsini/III/astro-ph> more vsiniIIIcor3.bbl 

\end{document}